\documentclass[aps,12pt,twocolumn,prl,nobibnotes,%
nofootinbib,superscriptaddress,noshowpacs,centertags]%
{revtex4-1}
\usepackage{graphicx}
\newcommand{\be}{\begin{eqnarray}}
\newcommand{\ee}{\end{eqnarray}}

\begin{document}
\title{CMB component separation in the pixel domain}

\author{A. Doroshkevich}
\address{Astro Space Center of Lebedev Physical
           Institute of  Russian Academy of Sciences,
                        117997 Moscow,  Russia\\}
\author{O. Verkhodanov}
\address{Special astrophysical observatory, Nizhnij Arkhyz,
Karachaj-Cherkesia}
\date{\today}

\begin{abstract}
We show that the popular ILC approach is unstable in respect to the
division of the sample of map pixels to the set of ``homogeneous''
subsamples. For suitable choice of such subsamples we can obtain the
restored CMB signal with amplitudes ranged from zero to the
amplitudes of the observed signal. We propose approach which allows
us to obtain reasonable estimates of $C_\ell$ at $\ell\leq 30$ and
similar to WMAP $C_\ell$ for larger $\ell$. With this approach we
reduce some anomalies of the WMAP results. In particular, our
estimate of the quadrupole is well consistent to theoretical one,
the effect of the ``axis of evil'' is suppressed and the symmetry of
the north and south galactic hemispheres increases. This results can
change estimates of quadrupole polarization and the redshift of
reionization of the Universe. We propose also new simple approach
which can improve WMAP estimates of high $\ell$ power spectrum.
\end{abstract}

\maketitle

\section{Introduction}

During last years fundamental results are obtained with the analysis
of  fluctuations of relic radiation
\cite{wmapresults,wmapfg,wmappara,wmap3ytem,wmap3ycos,
wmap5ytem,wmap5ycos,wmap7ytem,wmap7ycos} observed by WMAP mission.
Key problem of such analysis is the cosmic microwave background
(CMB) component separation from the Galactic foregrounds in the
pixel domain. Several approaches were used to separate CMB from the
observed signal. They are internal linear combination (ILC) and
maxima entropy methods \cite{wmap3ytem, efst04}, the blind and
Wiener filtering methods \cite{teg_efst, toh}, harmonic ILC
\cite{knc}, fast independent component analysis (FASTICA)
\cite{fastica} etc. Among these approaches the ILC method is very
convenient because in fact it requires minimal additional
assumptions in respect to the separated signals. Detailed discussion
of the ILC approach with many corrections can be found in
\cite{wmap3ytem, efst04}. The instability of the low multipoles
reconstruction with the ILC method owing to the correlation between
the CMB and foregrounds was discussed in \cite{instab_nas}. Recently
some problems arising with the ILC method were discussed in
\cite{ilc09}.

In Planck review \cite{planck_separation}, there is considered the
final component separation pipeline for the Planck mission, which
involves a combination of methods and iterations between processing
steps targeted at different objectives such as diffuse component
separation, spectral estimation, and compact source extraction

At the same time some anomalies in results of WMAP team are widely
discussed. Among other these are the small amplitude of quadrupole
component, unexpected correlations between components with
$\ell=2\,\&\,3$ (``axis of evil''), noticeable asymmetry between
north and south galactic hemispheres, existence of few deep walls in
the CMB map etc. Final step with these discussions is paper
\cite{bennet10} where all these anomalies are explained as random
fluctuations.

Special problem is the analysis inhomogeneous map for which the
amplitudes of foregrounds strongly vary over the map. In this case
the analysis becomes more complex and as is described in
\cite{wmap3ytem} it includes the division of the map to set compact
more homogeneous regions for which the component separation is
performed independently. However in \cite{wmap3ytem} the choice of
12 such regions is not uniquely determined. Different definitions of
"homogeneity" of selected subsamples are possible what leads to
different final estimates of the CMB map and $C_\ell$.

In this paper we show that the ILC method is unstable in respect to
the definition of "homogeneous" regions. As is shown below different
criteria of homogeneity and corresponding division of the full
sample of map pixels to set of "homogeneous" subsamples leads to
different CMB maps and even different $C_\ell$. Thus, for suitable
procedure we can obtain the CMB signal in wide range of its
amplitude. In fact these amplitudes can vary from zero to the
amplitude of observed signal.

In Section 2 we represent four different procedure which can be used
for the division of the map pixels in the set of 'homogeneous'
subsamples with analytical and numerical estimates of efficiency CMB
component separation. In Section 3 we apply our "best" approach to
the observed Q and V channels of WMAP and show that we can suppress
some of the anomalies noted above. Sec. 4 includes the summary of
our results and discussion of methodical problems. In particular, we
propose new approach for the analysis of high $\ell$ power spectrum
which can improve now available results.

\section{Separation of the CMB signal with ILC approach}

\subsection{The ILC approach}

The observed map is builded as a set of pixels each of which
contains combination $S(\theta_i)$ of the CMB signal $C(\theta_i)$
and the foreground $F(\theta_i)$. If we have maps at two different
frequencies then we can write
\begin{eqnarray}
S_1(\theta_i)=C(\theta_i)+F_1(\theta_i)\,,
\label{basic}
\end{eqnarray}
\[
S_2(\theta_i)=C(\theta_i)+F_2(\theta_i),
\]
and we like to perform the linear extraction of the CMB signal as
follows
\[
C(\theta_i)=\alpha S_1(\theta_i)+(1-\alpha)S_2(\theta_i)
\]
\begin{eqnarray}
 =S_2(\theta_i)+\alpha [S_1(\theta_i)-S_2(\theta_i)]\,,
\label{linear}
\end{eqnarray}
The general expression for $\alpha$ determined by the condition of
minimal dispersion of cleaned map is \be \alpha=-\langle
Q_2Q_{12}\rangle/\langle Q_{12}^2\rangle\,, \label{com} \ee
\[
\sigma_C^2=\langle C^2\rangle-\langle C\rangle^2= \langle
Q_2^2\rangle -
\langle Q_2Q_{12}\rangle^2/\langle Q_{12}^2\rangle\,.
\] Here
\[
Q_1(\theta_i)=S_1(\theta_i)-\langle S_1\rangle,\quad
Q_2(\theta_i)=S_2(\theta_i)-\langle S_2\rangle\,,
\]
\[
Q_{12}(\theta_i)=Q_1(\theta_i)-Q_2(\theta_i),\quad \langle
Q_1\rangle=\langle Q_2\rangle=0\,.
\]
and $\langle \rangle$ means the averaging over the considered
subsample of pixels.

However, as is seen from (\ref{basic} \& \ref{linear}),
\begin{eqnarray}
\alpha\langle (1-F_1/F_2)\rangle=1,\quad \alpha=\alpha_f=-(1-\langle
F_1/F_2\rangle)^{-1} \label{alph-f}
\end{eqnarray}
where in accordance with the main ideas of the approach we consider
$\alpha$ as a constant.

Relation (\ref{alph-f}) points out the best value of the parameter
of separation $\alpha=\alpha_f$. This value depends upon the ratio
$F_1/F_2$ and the scatter of $\alpha$ is determined by the scatter
of this ratio for the subsample used. Moreover, two values of the
parameter of separation, $\alpha$ (\ref{com}) and $\alpha_f$
(\ref{alph-f}), are different and this difference decreases for
decreased scatter of ratio $F_1/F_2$. This means that in order to
improve the separation we must divide the full sample of pixels to
set of more homogeneous subsamples using the distribution of ratios
$F_1\theta_i)/F_2(\theta_i)$. After the component separation within
these subsamples we get set of cleaned pixels sum of which forms the
cleaned map and allows to perform further analysis of this map with
better precision. Example of such component separation is considered
below (model 1).

However, such approach cannot be used in practice when the
foregrounds are a priory unknown and for the component separation we
would have to use criteria expressed through the observed signals.
As we show below the cleaned map strongly depends upon these
criteria.

In the further analysis we consider the pixels as independent ones
and ignore the possible correlations of the signal amplitude in the
neighboring pixels. The inclusion of such correlations allows to
improve the component separation but makes the procedure of
separation more complex.

As demonstration of these statements we consider below both
analytically and numerically four models of map division on
"homogeneous" \,subsamples prepared with various definitions of
"homogeneity". We determine the ``homogeneous'' subsamples in
respect to the function $G$ of amplitudes of signals
\[
G_i=G(\theta_i)=G(S_1(\theta_i), S_2(\theta_i))
\]
The $i^{th}$ bin contains $K_i$ pixels for which we have
\be
i\leq G_i/ \Delta\leq i+1
\label{Dlt}
\ee
where $\Delta$ is a given common width of the bins.
The bin center is the mean amplitude of the function $G_{ik}$
\be
\langle G_i\rangle=\sum_{k=1}^{K_i} G_{ik}/K_i\,,
\ee
By the way for all bins we have the symmetric distribution of
functions $G_{ik}$ with
\[
|\delta_k|=|G_{ik}-\langle G_i\rangle|\leq\Delta,\quad
\langle\delta_i\rangle=\langle G_i-\langle G_i\rangle \rangle\equiv
0\,.
\]

For each subsample we obtain $\alpha_i$ according to the standard
relation (\ref{com}) and get the CMB signal, $C(\theta_{ik})$ for
each pixel of considered subsample with relation \ref{linear}.

In main this approach is similar to that used in \cite{wmap3ytem}
in order to take into account the inhomogeneities of the foreground.
 However, their selection of 12 pixel subsamples differs from ones
discussed below. Our analysis confirms that the correct result can
be obtained only for the known a priory foregrounds. In all other
cases we can obtain the approximate estimate of the CMB signal only.
But deviations between the input and restored CMB signals depend
upon the criteria homogeneity and decreases for less $\Delta$. For
larger $\Delta$ all approaches give comparable results.

\subsection{Four models of separation of the CMB signal}

The theoretical consideration reveals the main influences of the
selection criteria but real estimates of quality of separation can
be found with simulations only. To test the various methods of
component separation we generate the CMB signals with the standard
power spectrum and Gaussian distribution of amplitudes, using the
foregrounds from WMAP \cite{wmap7yfgd} we transform the generated
CMB signals to observed ones and separate the CMB signals with
various approaches. The final estimates of precision achieved for
the full map relate to the comparison of introduced and restored
$C_\ell$.

\subsubsection{model 1}

Let us consider the set of subsamples with
\be
G(\theta_i)=F_1(\theta_i)/F_2(\theta_i)=1+\beta+\delta(\theta_i)\,,
\label{ff}
\ee
\[
\langle G\rangle=1+\beta,\quad |\delta|\leq \Delta\,,
\]
\[
F_1=F_2(1+\beta+\delta_i),\quad \langle Q_{12}\rangle=\beta\langle
F_2\rangle+\langle F_{2} \delta\rangle\,,
\]
 Here $1+\beta$ is the center of the subsample
and $\delta_i=\delta(\theta_i)$ characterizes the (small) random
scatter of the pixel amplitude in respect of the central point
($\langle\delta\rangle=0$)\,.

For such subsample we get
\[
\alpha=-\frac{1+o(\delta)}{\beta+o(\delta)},\quad
\alpha_f=-\frac{1}{\beta}\,,
\]
and for $\delta\rightarrow 0$ we have $\alpha\rightarrow
\alpha_f=-1/\beta$,
 \be
C(\theta_i)=C(\theta_i)+\Delta_C(\theta_i)\,, \label{cf} \ee
\[
\Delta_C(\theta_i)=F_2(\theta_i)\frac{\langle
F_2\delta\rangle/\langle
F_2\rangle+\delta(\theta_i)}{\beta+\delta(\theta_i)}\propto
\delta\,.
\]
As is seen from this relation
\be \Delta_C(\theta_i)\rightarrow
0\quad {\rm for}\quad \delta\rightarrow 0 \label{err1} \ee

For such choice of the pixel subsamples we get accurate component
separation precision of which depends upon the bin size, $\Delta$,
and increases for smaller $\Delta$. Numerical simulations confirm
this conclusion.

\subsubsection{model 2}

Let us consider the set of the pixel subsamples with
\be
G_i=S_1(\theta_i)/S_2(\theta_i)=1+\beta+\delta(\theta_i)
 \label{ss}
\ee where again $1+\beta$ is the center of the subsample and
$\delta(\theta_i)$ characterizes the (small) random scatter of the
pixel amplitude in respect of the central point
($\langle\delta\rangle=0$, $|\delta|\leq \Delta$). In the case
\[
S_1(\theta_i)-S_2(\theta_i)=S_2(\theta_i)(\beta+\delta_i),\quad
Q_2(\theta_i)=S_2(\theta_i)-\langle S_2\rangle\,,
\]
\be
Q_{12}(\theta_i)=\beta Q_2(\theta_i)+S_2(\theta_i)
\delta(\theta_i)-\langle S_2\delta\rangle
\label{det}
\ee
\[
\langle Q_{12}^2\rangle=\beta^2\langle Q_2^2\rangle+ 2\beta\langle
\delta S_2Q_2\rangle+o(\delta^2)\,,
\]
\[
\langle Q_2Q_{12}\rangle=\beta\langle Q_2^2\rangle+ \langle\delta
S_2Q_2\rangle,\quad \alpha\approx -1/\beta+o(\delta)\,.
\]

Therefore, \be C(\theta_i)=S_2(\theta_i)\frac{\langle\delta
S_2Q_2\rangle+ o(\delta^2)}{\beta \langle
Q_2^2\rangle+o(\delta)}\propto o(\delta)\,, \label{c2} \ee
\[
\sigma_C^2=\langle
Q_2^2\rangle\left[1-\frac{1+o_1(\delta)}{1+
o_2(\delta)}\right]\propto o(\delta)\,.
\]
Thus, we see that $C(\theta_i)\propto\Delta$,
$\sigma_C^2\propto\Delta$, and for $\Delta\rightarrow 0$ we have
~$C(\theta_i)\rightarrow 0,~~\sigma_C^2\rightarrow 0$. For such
pixel subsamples we get the extremal result  -- the signal CMB equal
zero. The same result can be obtained for an arbitrary function
$G=G(S_1/S_2)$. Numerical models confirm this tendencies.

\subsubsection{model 3}

Let us consider the set of pixel subsamples with
\be
G_i=S_1(\theta_i)=S_0[1+\delta(\theta_i)]\,,
\label{s0}
\ee
\[
Q_{12}=S_0\delta(\theta_i) -Q_2\,.
\]
Here $S_0$ is the center of the subsample and $\delta(\theta_i)$
characterizes the (small) random scatter of the pixel amplitude in
respect of the central point ($\langle\delta\rangle=0$,
$|\delta|\leq\Delta/S_0$). In the case \[ \langle
Q_{12}^2\rangle=\langle Q_2^2\rangle - 2S_0\langle\delta
S_2\rangle+S_0^2\langle\delta^2\rangle\,, \]
\[
\langle Q_2Q_{12}\rangle=-\langle Q_2^2\rangle+ S_0\langle\delta
Q_2\rangle,\quad \alpha=\frac{1-o_1(\delta)}{1-o_2(\delta)}
\]
\be C(\theta_i)=S_2(\theta_i)(1-\alpha)+\alpha
S_0[1+\delta(\theta_i)]\,. \label{s03} \ee

Thus, for $\delta\rightarrow 0$ we get
\be
\alpha\rightarrow 1\quad
C(\theta_i)\rightarrow S_0,\quad \sigma_C^2\rightarrow 0\,.
\label{ex3}
\ee

For such choice of the function $G_i$ (\ref{s0}) we get unexpected
result  - for small $\Delta\rightarrow 0$ the signal CMB is equal to
$S_1=S_0$. Numerical simulations confirm these tendencies and as is
seen from the Table 1 for small $\Delta$ the selected signal $C$ is
quite close to the input one $S_1$ and strongly differs from $S_2$.
For larger $\Delta$ this difference disappears.

\begin{table}[!th]
\caption{Two examples of the reconstruction of the CMB signal with
the model 3 (arbitrary units)} \label{tbl1}
\begin{tabular}{lrlll}
\hline
   $\Delta$&Npixels&$~~~~\langle S_1\rangle$&~~~~$\langle S_2\rangle$&
   ~~~~$\langle C\rangle$\cr
\hline 0.2mK&256129&$8.7\pm 5.4$&$~~~~5.6\pm 5.3$&$3.3\pm 6.3$\cr
0.002mK&2835&$1.0\pm 0.57$&$-20.~\pm 17.$&$0.4\pm 0.8$\cr
0.002mK&2922&$3.0\pm 0.6$&$-18.~\pm 18.$&$1.3\pm 1.6$ \cr \hline
\end{tabular}
\end{table}

\begin{figure}[!th]
\includegraphics[width=0.63\linewidth,height=1.\linewidth,angle=-90
]{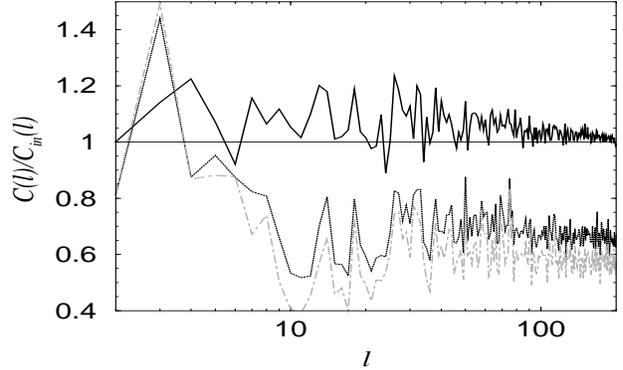}
\caption{For the model 3 the reconstruction of modeling CMB signal
for the foregrounds in Q and V bands. Results are normalized on the
input signal and are shown for $\Delta=2, 0.2\,\&\,0.002 mK$ (solid,
dashed and dot-dashed lines). } \label{fig_mod3}
\end{figure}

For $\Delta=2,\,0.2\,\&\,0.002mK$ reconstruction of the modeling CMB
signal with foregrounds in Q and V bands are presented in
Fig.\ref{fig_mod3}. It is interesting that the best reconstruction
is obtained for the larger $\Delta$ and for restored and input
signals the ratio $C_\ell/C_{in}$ decreases with $\Delta$.

\subsubsection{model 4}

Let us consider the set of pixel subsamples with \be
G_i=S_1(\theta_i)-S_2(\theta_i)=F_1(\theta_i)-F_2(\theta_i)=
\beta(1+\delta_i) \label{s4} \ee
\[
Q_{12}=\beta\delta(\theta_i)\quad \langle\delta\rangle=0,\quad
|\delta_i|\leq \Delta/\beta\,.
\]
Here $\beta$ is the center of the subsample and $\delta_i=
\delta(\theta_i)$ characterizes the (small) random scatter of the
pixel amplitude in respect of the central point. As in the model 1,
so determined function $G_i$ depends upon the foregrounds only what
is some advantage of this approach. In the case
\[
\langle Q_2Q_{12}\rangle=\beta\langle S_2\delta\rangle,\quad \langle
Q_{12}^2\rangle=\beta^2\langle\delta^2\rangle,\quad
\alpha=-\frac{\langle\delta
S_2\rangle}{\beta\langle\delta^2\rangle}\,,
\]
\be C(\theta_i)=S_2(\theta_i)-[1+\delta(\theta_i)] \langle
S_2\delta\rangle/\langle\delta^2\rangle\,, \label{ss4} \ee
\[
\sigma_C^2=\langle Q^2_2\rangle-\langle S_2\delta\rangle^2/
\langle\delta^2\rangle
\]

For such choice of the function $G_i$ results depend upon the bin
size but even for $\delta\rightarrow 0$ they are not tend to real
CMB signal. In the case the choice of optimal $\Delta$ can be done
with simulations.

\begin{figure}[!th]
\includegraphics[width=0.63\linewidth,height=1.\linewidth,angle=-90
]{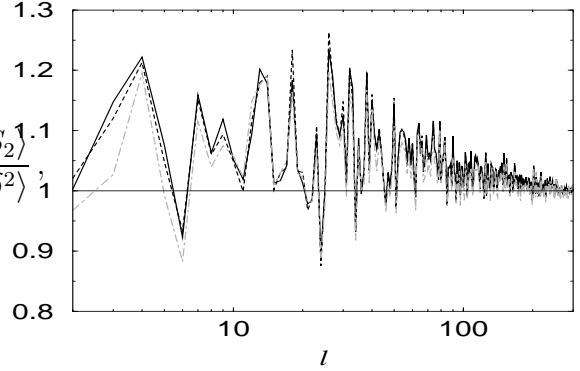}
\caption{For the model 4 the reconstruction of modeling CMB signal
for the foregrounds in Q and V bands. Results are normalized on the
amplitude of input signal and shown for $\Delta=2, 0.2\,\&\,0.002
mK$ (solid, dashed and dot - dashed lines). } \label{fig_mod4}
\end{figure}

Examples of such reconstruction of the input CMB signal with
$\Delta=2, \,0.2,\,\&\,0.002mK$ are presented in Fig.
\ref{fig_mod4}. As is seen from this figure reconstructed signal is
weakly sensitive to used small $\Delta$ and is oscillated around the
level $C_\ell/C_{in}\sim 1. - 1.1$.

\begin{figure}[!th]
\includegraphics[width=1.\linewidth,height=0.5\linewidth]{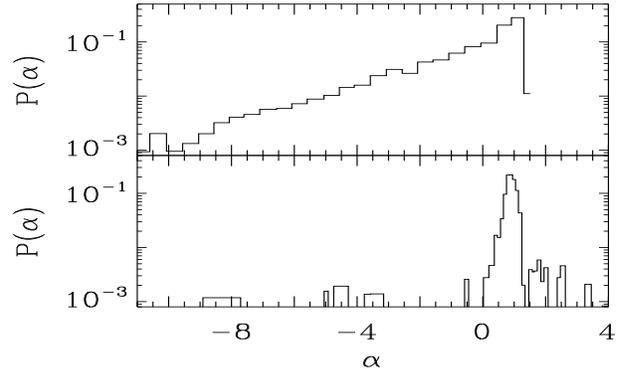}
\vspace{0.4cm} \caption{For $\Delta=0.02mK$ the distribution of the
fraction of pixels vs. $\alpha$ for model 3 (top panel) and model 4
(bottom panel) } \label{f-alpha}
\end{figure}

The difference between models 3 and 4 is illustrated by Fig.
\ref{f-alpha} where we see the probability distribution function
$P(\alpha)$ for fraction of pixels versus the separation coefficient
$\alpha$.

\section{Power spectrum from Q and V bands of WMAP }

\begin{figure}[!th]
\includegraphics[width=0.95\linewidth,height=0.45\linewidth,
]{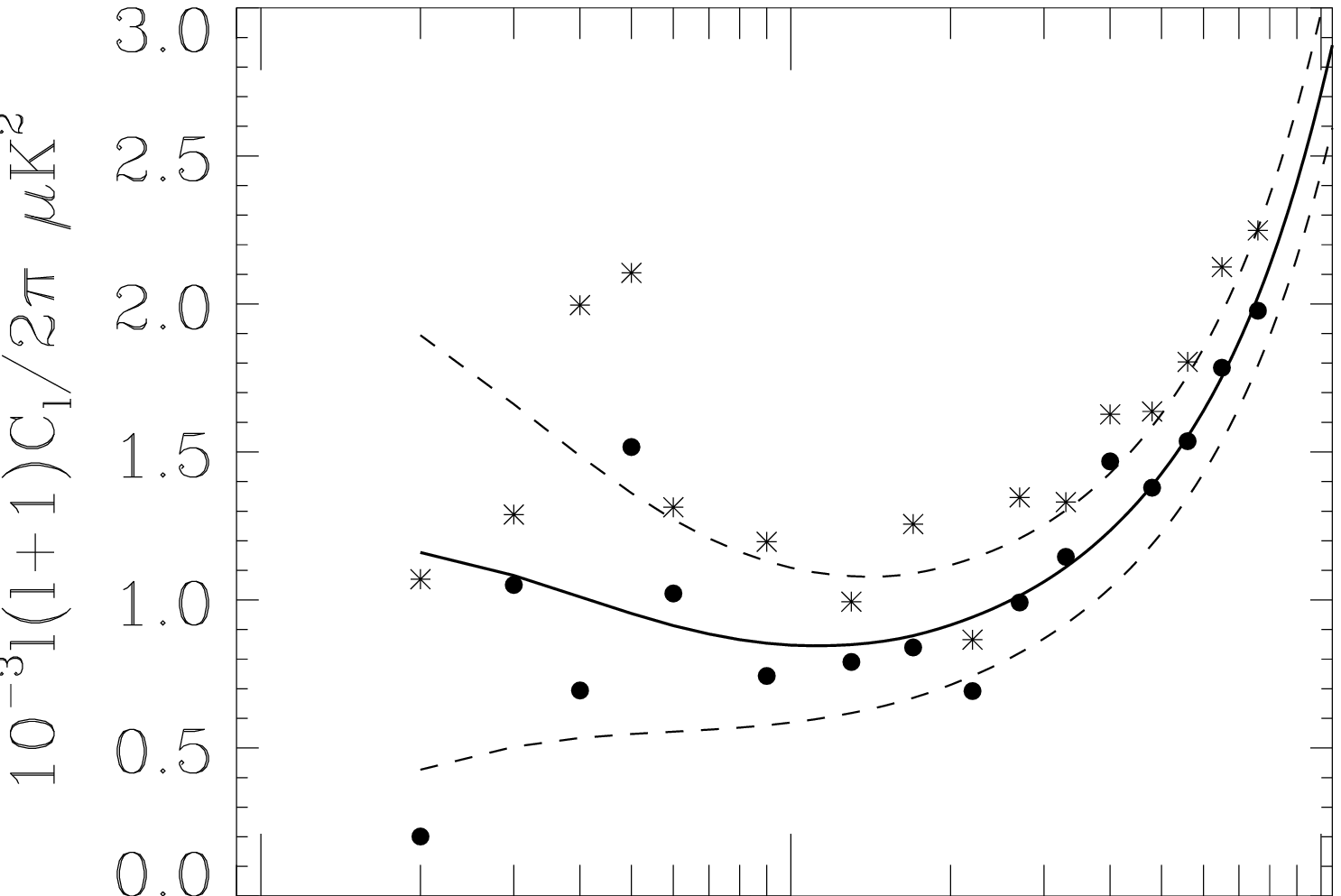}
\vspace{0.5cm}
 \caption{The $10^{-3}\Delta T^2$ for the WMAP data (points)
and obtained according to the method used for the model 4 (stars)
with $\Delta=20\mu K$. Solid and dashed lines show the theoretically
expected values and their scatter.} \label{fig_cmb}
\end{figure}

As was found in previous Section the best reconstruction of $C_\ell$
is possible with approach used in the model 4. Applying this
approach with $\Delta\leq 0.2mK$ for Q and V bands of the WMAP maps
we get $C_\ell$ which significantly differ from ones presented in
WMAP publications. In these cases we have from several tens to
several thousands of 'homogeneous' regions instead of 12 regions
used in WMAP analysis. These $\Delta T^2_\ell$ are plotted in Fig.
\ref{fig_cmb} and $a_{2m}$ are listed in Table \ref{tblII}. However,
for broad bins with $\Delta\geq 10mK$ our results become quite
similar to the WMAP ones.

\begin{table}[!th]
\caption{Amplitudes of quadrupole components in $\mu$ K for
$\Delta=2\mu K$} \label{tblII}
\begin{tabular}{lrr}
\hline
        &WMAP&Model 4\cr
\hline $a_{2, 0}$&11.48  &-65.2  \cr $a_{2, 1}$&-0.05  &-13.8  \cr
$a_{2,-1}$&4.86   &9.0  \cr $a_{2,2}$ &-14.41 &-17.3  \cr $a_{2,-2}$
&-18.80&-11.0  \cr \hline
\end{tabular}
\end{table}
With $a_{2m}$ listed in Table \ref{tblII} we get for the quadrupole
\be \Delta T^2_Q\approx 1070\mu K^2\,, \label{c2} \ee what is close
to theoretical expectations \cite{wmappara} \be \Delta
T^2_{th}\approx 1250 \mu K^2\,. \label{th} \ee and exceeds estimate
obtained by WMAP team \cite{wmap3ytem} \be \Delta T^2_Q\approx 249
\mu K^2\,. \label{wmap} \ee

As is well known, the five quadrupole coefficients are equivalent to
the components of a symmetric traceless tensor.
For the principle values and orientation of tensor
axes for the 3 years WMAP quadrupole we have \cite{dd07}
\[
\lambda_1=~~27.1 \mu K,\quad (l,b)=(-0.8^\circ\pm 13^\circ,\,\,
63.3^\circ\pm 1^\circ)\,,
\]
\be \lambda_2=~~12.9 \mu K,\,\, (l,b)=(15.5^\circ\pm 3^\circ,\,\,\,
25.8^\circ\pm 1.2^\circ),
\label{ilc3} \ee
\[
\lambda_3=-40~~\mu K,\quad (l,b)=(-77.6^\circ\pm 5^\circ,\,\,\,
6.5^\circ\pm 4^\circ)\,,
\]

In contrast, for our parameters of quadrupole we get
\[
\lambda_1=~~68.3 \mu K,\quad (l,b)=(-75^\circ,\,\,\,~~9.1^\circ)\,,
\]
\be
\quad~~\lambda_2=~~12.0 \mu K,\quad (l,b)=(13.1^\circ,\,\,\,
-8.7^\circ),
\label{ilc33}
\ee
\[
\lambda_3=-80.4\mu K,\quad ~~(l,b)=(60.^\circ,\,\,\,~~77.4^\circ)\,,
\]
with
\[
\Delta T^2=-\frac{3}{5\pi}(\lambda_1\lambda_2+\lambda_1\lambda_3+
\lambda_2\lambda_3)=1070 \mu K^2\,.
\]
The orientations (\ref{ilc33}) differ from both the dipole direction
\[
(l,b)_D=(-96^\circ,48^\circ)\,,
\]
and from orientations (\ref{ilc3}).

As is seen from Fig. \ref{fig_cmb} the most serious differences are
found for $\ell=2,\,4,$ and for even $\ell\leq 30$. For these even
$\ell$ our estimates of $C_\ell$ exceed ones obtained by WMAP by a
factor of $\sim 1.5$ what emphasizes the symmetry of the CMB signal
in north and south hemispheres. For $\ell\geq 30$ the difference
becomes small. It is interesting that for odd $\ell$ deviations from
WMAP results regularly do not exceed 10\%. The random scatter of the
method depends upon the bin size used, $\Delta$, but not exceed
$\sim 10 - 15\%$ what does not distort essentially our estimates of
the power spectrum.

These results noticeably change low $\ell$ part of the power
spectrum and significantly suppress the effect of ``axis of
evil''. However, they do not distort strongly main conclusions
of WMAP which are weakly depend upon this part of the power
spectrum.

Let us emphasis only that new estimate of $C_2$ can noticeably
changes the estimates of the quadrupole polarization and,
therefore, the redshift of reionization.

\section{Summary and discussion}

In this paper we show that the separation of foregrounds and the CMB
signal with the ILC method strongly depends upon the choice of
'homogeneous' subsamples of pixels. For foregrounds presented in
WMAP papers our more stable estimates of the CMB fluctuations are
obtained for the selection criteria used in the model 4. Theoretical
consideration (\ref{s4}) shows that with this approach we cannot
perform the very high precision cleaning. However, numerical
analysis demonstrates that for suitable choice of the bin size,
$\Delta$, the precision $\sigma\approx 10\%$ can be achieved. It can
be expected that the application of refined technique developed by
WMAP team will allow to decrease the errors up to values presented
in \cite{wmap3ytem}.

\subsection{Main results}

The best results are obtained for the frequency channels Q \& V and
are presented in Fig. \ref{fig_cmb}. Main results of our analysis
can be summarized as follow:
\begin{enumerate}
\item{} The measured amplitude of quarupole  is more than that given
by WMAP by a factor of 2.1 what eliminates disagreement between the
theoretically expected and measured values.
\item{} The coordinates of the quadrupole are changed while our
estimates of the octupole remain the same as in WMAP. This fact
substantially reduces the effect of "axes of evil".
\item{}All even $C_\ell$ with $4\leq\ell\leq 20$ are more then those
given by WMAP by a factor of $\approx 1.5 - 2$ what emphasizes the
symmetry of the CMB signal in north and south hemispheres.
\item{} Deviations of odd $C_\ell$ from that given by WMAP do not
exceed a factor of 1.2 - 1.3 .
\item{} At $\ell\geq 30$ deviations of our estimates from the WMAP
data do net exceed $\approx 5\%$.
\item{} At $\ell\leq 20$ the expected error of measured $C_\ell$
is $\sim 10\%$.
\end{enumerate}

These results indicate that the main conclusions of the WMAP team
remain correct. However, the change of the large scale
characteristics leads to the moderate change of estimates of
$\sigma_8$ and especially the estimates of low $\ell$ polarization
and, therefore, the redshift of reionization of the Universe. These
corrections could be important for analysis of the epoch of
reionization and formation of earlier galaxies.

The further more detailed analysis of possible divisions of the full
sample of pixels to the 'homogeneous' subsamples can find more
effective methods of subsample selection than that used in the
paper. In particular, the account of correlation of the signal
amplitude in neighboring pixels can improve the quality of the
cleaned map of the CMB signal.

Of course, this approach can be extended for the three and more
frequency channels.

\subsection{Methodical comments}

The considered models allow us to obtain some inferences related
to the method of linear component separation. Thus, we see that:
\begin{enumerate}
\item{} The method of linear component separation is unstable and
the resulting CMB map strongly depends upon criteria homogeneity
used for the selection of the set of subsample under consideration.
\item{} The best separation is possible with using the foreground
measurements (model 1). However, such approach is of no concern for
a practice as we do not know a priori the foregrounds.
\item{} Models 2 and 3 demonstrate that with a suitable choice of the
selection criteria we can obtain arbitrary estimates for the CMB
signal.
\item{} Reasonable estimates of the CMB signal can be obtained
with the selection criteria used in the model 4. However even in
this model the CMB signal can be found with errors which depend upon
the bin size $\Delta$ (\ref{Dlt}) used for the subsample selection.
\item{} Comparison of theoretical estimates of $\sigma_C$ for models
3 and 4 with numerical estimates of $C_\ell$ shows that sometime the
former ones do not characterize adequately the final precision
achieved.
\end{enumerate}

It can be expected that final results depend upon the actual
foreground. This inference is confirmed by comparison results
obtained for various pairs of frequency channels.

Let us note that further cleaning can be performed by recurrent
comparison of the cleaned maps obtained for two pairs of
frequencies. With the WMAP data we cannot test this approach as the
quality of maps obtained for QV channels significantly exceed the
quality of maps found for other pairs of frequency channels.
However, for many channels of the PLANCK mission such approach
becomes useful.

\subsection{Estimates of the high $\ell$ power spectrum}

As is well known for the real maps of the CMB with the finite
number of pixels the determination of the power spectrum for
larger $\ell$ is complex because the polar regions with relatively
small number of pixels along the azimuthal coordinate cannot be used.
By the way at high $\ell$ we would have to analyze the noisy regions
in the vicinity of equator what decreases the precision achieved.

\begin{figure}[!th]
\includegraphics[width=0.7\linewidth,height=0.4\linewidth,
]{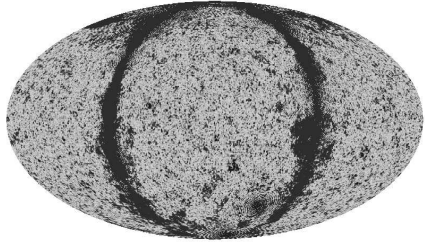}
\caption{The CMB map for the Q channel after rotation of the
coordinate system.} \label{rot_map}
\end{figure}

To decrease the influence of the noisy galactic equator we can use
the simple procedure what is change of the map orientation. Indeed,
if we will build the map in coordinate system with the galactic
equator situated along some map meridian then we will have less
noisy pixels situated along the map equator while some of the noisy
pixels will be shifted to polar regions. Example of such map is
presented in Fig. \ref{rot_map}.

Of course, such approach requires preparation of two different
maps one of which have the ordinary orientation and is used for
the analysis of the low $\ell$ part of power spectrum while
second one with the orthogonal orientation can be used for
analysis of high $\ell$ components of the power spectrum.

This approach seems to be quite effective but it must be tested
with real repixelized maps.

\section{Acknowledgments}
We thank the NASA for making available the NASA Legacy Archive, from
where we adopted the WMAP data. We are also grateful to the authors
of the HEALPix\footnote{\tt http://healpix.jpl.nasa.gov/}
\cite{healpix} package, which we used to transform the WMAP7 maps
into the coefficients $a_{\ell m}$. This work made use of the GLESP
\footnote{\tt http://www.glesp.nbi.dk} \cite{glesp,glesp1} package
for the further analysis of the CMB data on the sphere. This paper
was supported  in part by Russian Foundation for Basic research
grant Nr. 08-02-00159 and Nr. 09-026-12163, and Ministry of
education Nr. 1336. O.V.V. also acknowledges partial support from
the "Dynasty" Foundation.

\def\apj{Astrophys. J}
\def\apjl{Astrophys. J. Lett.}
\def\apjs{Astrophys. J. Supp.}
\def\ijmpd{Int. J. Mod. Phys. D}
\def\mn{Mon. Not. R. Astron. Soc.}
\def\nature{ Nature}
\def\aa{Astro. \& Astrophys.}
\def\prl{Phys.\ Rev.\ Lett.}
\def\prd{Phys.\ Rev.\ D}
\def\pr{Phys.\ Rep.}

\end{document}